\documentclass[12pt]{iopart}

\usepackage{iopams}  
\usepackage{graphicx}
\usepackage[breaklinks]{hyperref}
\usepackage{cite}
\begin{document}

\title{Synchronization gauge field, standing waves and one-way-speed of light}

\author{Arunava Bhadra$^{1,a}$, Abhishek Chakraborty$^{1,b}$, Souvik Ghose$^{2,c}$, Biplab Raychaudhuri$^{3,d}$ }

\address{$^1$High Energy Cosmic Research Center, University of North Bengal, Siliguri, Darjeeling, West Bengal, India, 734013}
\ead{$^a$aru\_bhadra@yahoo.com}
\vspace{10pt}
\ead{$^b$abhichak7@gmail.com}
\vspace{10pt}
\address{$^2$Department of Physics, Harish-Chandra Research Institute, Allahabad (Prayagraj), UP, India, 211019}
\ead{$^c$souvikghose@hri.res.in}
\vspace{10pt}
\address{$^3$Department of Physics, Visva-Bharati University, Santiniketan, West Bengal, India, 731235}
\ead{$^d$biplabphy@visva-bharati.ac.in}
\vspace{10pt}

\begin{indented}
\item[]October 2023
\end{indented}

\begin{abstract}
The absolute nature of many fundamental predictions of the theory of special relativity, including the   relativity of simultaneity, has been questioned in the literature owing to the choice of distant clock synchronization process in the theory. Here we discuss the consequences of Anderson-Vetharaniam-Stedman (AVS) conventionality synchronization gauge, which reflects the choice of synchronization convention, on the standing wave observable. We found that although the position of the node(s) is gauge invariant and remain the same as in the standard case of the stationary wave formation following the Einstein synchronization, the anti-node(s) becomes a gauge dependent (conventional) element and the resulting wave travels between two nodes, contrary to the experimental observation. The experimental detection of standing wave substantiates that the one-way velocity is equal to the round-trip velocity implying the uniqueness of the Einstein synchronization convention. The present analysis thus eliminates the (unphysical) synchronization gauge freedom of special relativity.    
\end{abstract}

\noindent{\it Keywords}: one way speed,  distant clock synchronization, standing waves
%\maketitle

\section{Introduction}

The standard description of the Special Theory of Relativity (STR) is based on the Lorentz transformation equations that relate two inertial frames of references (IRF) moving with respect to each other. However, several authors have pointed out that the Lorentz transformation is not the unique transformation equations derivable from the basic postulates of STR \cite{anderson1998, Selleri97, Rizzi05} and the general transformation equations are characterized by a vector field that (apparently) does not have any effect on any observable and therefore appears as a gauge field. %A particular choice of such gauge field leads to the Lorentz transformation. 
The physical origin of the stated gauge field is the freedom of choice of the synchronization of distant clocks, and hence it is called as the synchronization gauge field. 

A (claimed) consequence of the synchronization gauge is that the isotropy of one-way speed (OWS) of light (in fact any signal) is not absolute, it rests on the choice of the gauge field \cite{anderson1998}. Note that while developing STR Einstein stipulated the equality of the one-way speed (OWS) of light with its round-trip speed \cite{einstein1922meaning,einsteinrelativity} which fixes the gauge field. Because of its definitional nature, many authors consider Einstein's stipulation as a mere convention \cite{reichenbach1958, grunbaum2012, anderson1998} (often called conventionality of (distant) simultaneity thesis). According to the conventionality thesis, all the derived expressions based on the isotropy of OWS of light, for instance, the Lorentz transformation, have conventional ingredient and as a result the observational effects followed from such expressions, such as the relativity of simultaneity, is not absolute. There were a few claims for experimental verification of isotropy of OWS \cite{Champeney65, Vessot80, Kaivola85, MacArthur86,riis1988test, krisher1990test, Hils90, will1992clock, greaves2009one}. But most of such claims, if not all, were subsequently countered in literature (\cite{anderson1998} and references therein) mainly on the ground that the observable of all such experiments are independent of synchronization gauge.

A gauge symmetry usually gives a deeper understanding of the underlying physics. In the present letter, we  revisit the well-known phenomenon of standing waves and examine whether the synchronization gauge has any effect on the observable of the standing wave. Besides the obvious reason for its formation by two plane waves of the same frequency and speed but moving in the opposite directions, our interest in the standing waves comes from its fundamental role in quantum mechanics. As we know, quantum mechanics describes particles and rigid bodies in terms of waves. In the framework of the field theory, the progressive (running) plane wave is employed in all calculation of Lorentz covariant S-matrix, which is associated with scattering processes under the perturbation approach. The S-matrix formalism is, however, valid for free particles. In the same field theoretical view, an extended object (bound state) consists of standing waves, which do not propagate and are formed as a result of interference between incident and reflected waves. So any additional degree of gauge freedom in the description of a standing wave may have its consequences in quantum field theory. 

The plan of the paper is as follows: In the next section, we would discuss the synchronization gauge in special relativity and efforts of testing isotropy of the speed of light experimentally. In section 3 we would analyze the superposition of incident and reflected waves under synchronization gauge transformation. The status of the experimental detection of the standing waves and their observables is discussed in section 4. Finally, we would conclude our results in section 5.  

\section{Synchronization gauge and measurement of isotropy of OWS of light}

Measuring the OWS requires two pre-synchronized clocks at two distant points while synchronization of distant clocks requires prior information on OWS of the synchronizing signal leading to a logical circularity which is the source of the synchronization gauge in STR. In the absence of any definite synchronization procedure, Einstein synchronization is usually employed for distant clocks synchronization where isotropic light propagation is assumed and corresponding transformation equations between two IRFs moving with a relative velocity v is the Lorentz transformation $X^{\prime \mu} = L^{\mu}_{\nu} X^{\nu}$ where $X^0 = ct$, $X^i=x^i$, and $L^0_0=L^1_1=\gamma$, $L^2_2=L^3_3=1$, $L^1_2=L^2_1=-\beta \gamma$ and the rest of the elements are zero, $\beta = v/c$ and $\gamma = (1-\beta^2)^{-1/2}$.     

Along with standard treatment, the possibility, in principle, of postulating an anisotropic structure has also been discussed extensively in various contexts. 
Mansouri and Sexl (MS) \cite{MS77} first introduced the synchronization gauge in STR under the so-called test theory formalism. The MS test theory is focused on kinematic aspects and space-time structure. Perhaps the most mathematically sound argument in favor of synchronization gauge (conventionality) in STR was advanced by Anderson {\it et al.} \cite{anderson1998} as described below. They proposed a coordinate transformation  

\begin{eqnarray}
\label{syn}
t^{\prime} = t- \boldsymbol{\xi}.\textbf{x}/c  \\ \nonumber
\textbf{x}^{\prime} = \textbf{x} 
\end{eqnarray}
%boldsymbol
where $\boldsymbol{\xi}$ is an arbitrary smooth vector field. Under the above transformation the locus of a light beam $x=\pm ct$ in the un-dashed, “Einstein synchronization” coordinates reduces to $x^{\prime} =\pm \frac{ct^{\prime}}{1 \mp \xi}$.

Thus the speed of light in each direction becomes $c_{\pm} = c/(1 \mp \xi )$. Since the physical laws remain invariant under coordinate transformation, $\mathbf{\xi}$ is the synchronization gauge field which is known as AVS synchronization gauge in the name of the authors. In the notation of Reichenbach $\xi=2 \epsilon -1$ so that these speeds are $\frac{c}{2\epsilon}$ and $\frac{c}{2(1-\epsilon)}$. Remember that the two-way speed is always $c$ as it can be measured with a single clock. For a particular gauge choice ($\xi = -\beta/c - \gamma^{-1} \epsilon(\beta)$ where $\epsilon(\beta)$ is the Selleri gauge parameter) AVS transformation becomes the general transformation between two IRFs as suggested by Selleri \cite{Selleri97}. 

\subsection{Experimental measurement of isotropy of OWS}
Despite the arguments of conventionalist that OWS is not measurable, several researchers that include a number of reputed scientists ventured for measuring OWS. The so-called test theories of Robertson \cite{Rob49} and that of Mansouri and Sexl \cite{MS77} were the main backbone for the purpose. 

A few noteworthy efforts/claims of (indirect) measurement of isotropy of OWS include the study of ionization effects on the hydrogen atoms by ultra-violet laser beam \cite{MacArthur86}, an improved Kennedy-Thorndike experiment under the MS formalism \cite{Hils90},  a rocket red-shift experiment \cite{Vessot80}, a M$\ddot{o}$ssbauer rotor experiment \cite{Champeney65}. The mentioned experiments claimed either the validity of the Lorentz transformation or the conformity of observed length contraction/time dilation as per the prediction of special relativity with Einstein synchronization.    

A notable experiment to examine the isotropy of OWS of light was performed by Riis et al \cite{riis1988test}. In their experiment, a beam of fast $^{20}Ne$ atoms were excited resonantly (via two-photon absorption) by two collinear anti-propagating laser beams of the same frequencies in the laboratory frame. The frequency of the laser beams was continually altered to maintain resonance. The frequency of the fast atom beam quantum transition was compared to the frequency of a stationary absorber as a function of rotation of the Earth. The Mansouri-Sexl transformation equations were used to relate the frequencies in the atomic rest frame with the frequencies of the laboratory frame. From the observational findings, the authors claimed that the level of anisotropy in the speed of light $\Delta c/c < 3 \times 10^{-9}$ \cite{riis1988test}. Kaivola {\it et al.}\cite{Kaivola85} performed a similar kind of experiment. 

A group of researchers from the Jet Propulsion Laboratory monitored the time of flight of signals along 29 km highly stable fibre optic cable (NASA deep space network) between two hydrogen maser clocks as a function of rotation of the Earth \cite{krisher1990test}. The authors claimed for isotropy of OWS of light from the non-observation of any diurnal variation in the phase of the incoming and outgoing signals at one end of the fibre cable. Will also analysed several of the experiments mentioned above in the framework of MS test theory \cite{will1992clock}.

The claim of validation of isotropy of light by above-stated experiments/analysis was contested by demonstrating that the observables of these experiments are independent of the choice of synchrony \cite{anderson1998} and references therein. The situation, thus, remains debatable. 

\section{Standing waves}
As is well known stationary (standing) waves are formed as a result of interference between incident and reflected waves. The underlying assumption is that the propagation velocity of the incident and reflected waves is the same. The vertical displacement of the traveling sinusoidal wave moving along $\pm$ x-direction can be expressed as

\begin{equation}
\label{eq:trav}
 y_{\pm}(x,t) = y_0 \sin(kx \mp  \omega t).                                                   
\end{equation}

where $y_0$, $\omega$ and $k$ have their usual meaning. %(in the above and subsequent equations the displacement $y_+$ and $y_-$ correspond respectively for the motion along +ve and -ve x directions. %are the amplitude of the wave, the angular frequency, and the wave number respectively. 
The resultant wave from the superposition of incident and reflected wave, which is the sum of the individual waves, is given by

\begin{equation}
\label{eq:std}
 y(x,t) = 2 y_0 \sin(kx)\; \cos (\omega t).                                                   
\end{equation}

As the x and t terms are separated, the resultant wave is no longer traveling. All particles of the wave will oscillate in time but have a spatial dependence that is stationary. At locations $x = 0, \lambda/2, \lambda, \, ...,$ called the nodes, the amplitude is always zero, whereas at locations $x =  \lambda/4, 3\lambda/4, 5\lambda/4, \, ...$, called the anti-nodes, the amplitude is maximum. The distance between two conjugative nodes or anti-nodes is $\lambda/2$.

Since the phase of a plane wave is Lorentz invariant, the form of the standing wave equation will not alter under the Lorentz boost. If the synchrony transformation is applied to the above progressive wave equation, the form of the equation will remain the same but the wavenumber (or velocity) in two directions will be different as $k_{\pm}=\frac{\omega}{c_{\pm}}$ where $ c_{\pm} = \frac{c}{1\pm \xi}$ (note that $\omega$ is independent of synchrony convention since frequency can be measured with a single clock).    

Applying the synchrony transformation (Eq.{\ref{syn}) to the standing wave equation (Eq.(\ref{eq:std})) with the direction of synchrony vector field along the x-direction, one gets
 
\begin{equation}
\label{eq:strw2}
y  (x, t) =  2 y_0  \sin \left( \frac{\omega x}{c}\right) \cos \left[-\xi\omega x/c +\omega t \right],
\end{equation}

The above equation describes a traveling wave where the position of the minimum remains the same as in the standard case (isotropic OWS) but the resulting wave travels between two minima. One may reach the above equation also by considering the superposition of the incident and reflected wave with different OWS. One would get back the standard standing wave equation from the above equation by considering the choice $\xi=0$, which is the Einstein synchronization. Clearly the anti-node points (in fact all other points except the nodes) are gauge dependent.  

In figures \ref{fig:sub-first} and \ref{fig:sub-second} the time variation of standing wave (Eq.(\ref{eq:std})) and the resultant wave (Eq.(\ref{eq:strw2})) (i.e. with two different choice of AVS gauge parameter ($\xi$)) are compared.

%%%%%%%%%%%%%  Figures %%%%%%%%%%%%%%%%%
%\centerline{\includegraphics[width=2.0in]{mplaf1}}
%\caption{A schematic illustration of dissociative recombination. The
%\protect\label{fig1}}
%\end{figure}
%\begin{subfigure}{.45\textwidth}
% \begin{subfigure}
% \centering
  % include first image
%  \includegraphics[width=.7\linewidth]{stn-1.eps}
\begin{figure}
%\begin{subfigure}
\centerline{\includegraphics[width=2.0in]{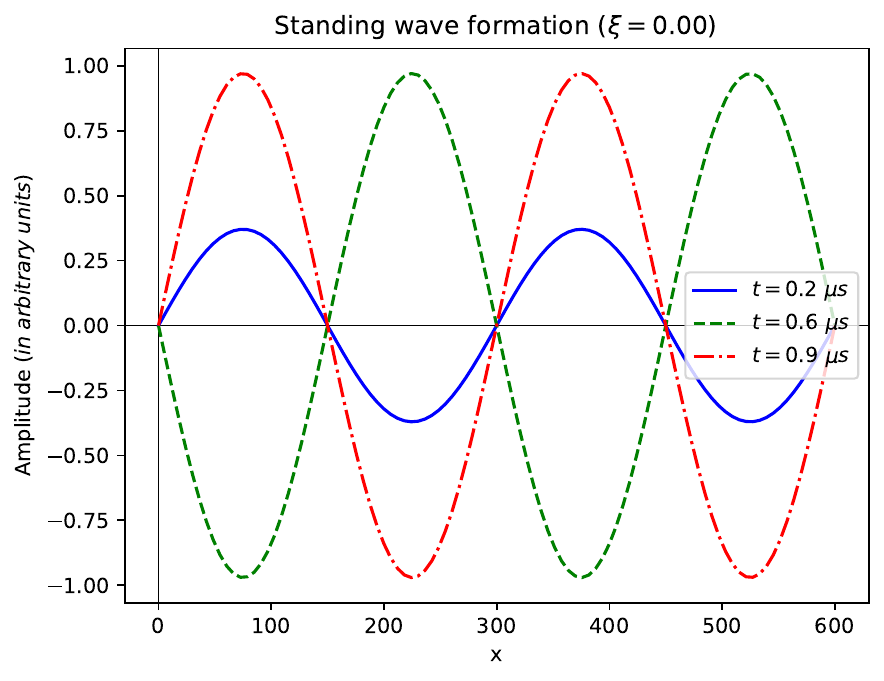}}
\centerline{\includegraphics[width=2.0in]{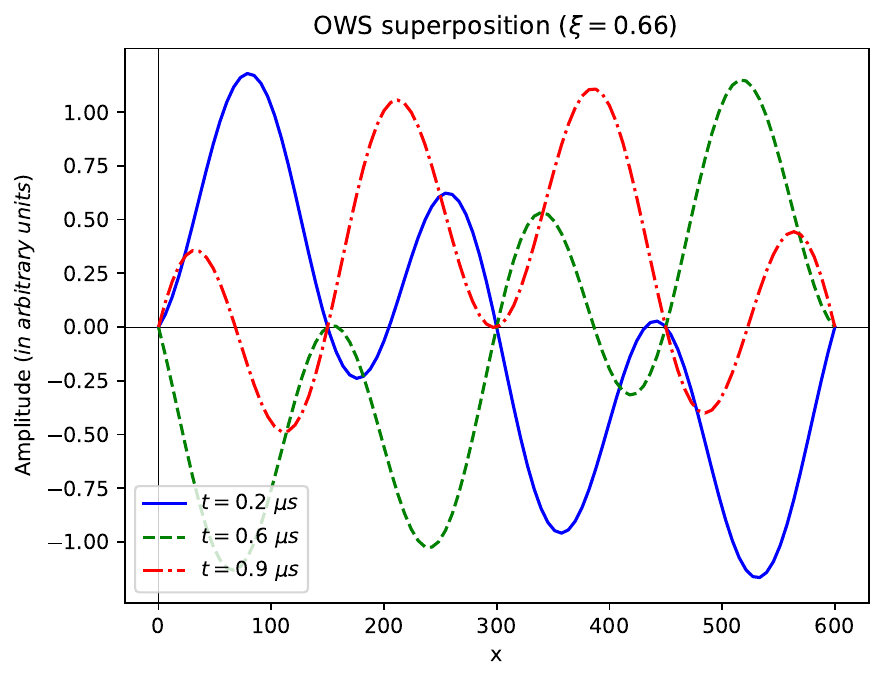}}

  \caption{A standard standing wave of a light signal in radio wave length ($\lambda=300$ m) at three diferent times (upper figure). The same for the resultant wave arriving after AVS gauge transformation with $\xi=0.66$ (lower figure) .}
\label{fig:sub-first}
%\end{subfigure}
%\begin{subfigure}
%  \centering
  % include second image
%	\begin{subfigure}
% \centerline{\includegraphics[width=2.0in]{stn-3.pdf}}
 %\includegraphics{stn-3.pdf}  
%  \caption{}
%  \label{fig:sub-second}
\end{figure}

\begin{figure}
% \centering
  % include third image
 % \includegraphics{stn-5.pdf}  
\centerline{\includegraphics[width=2.0in]{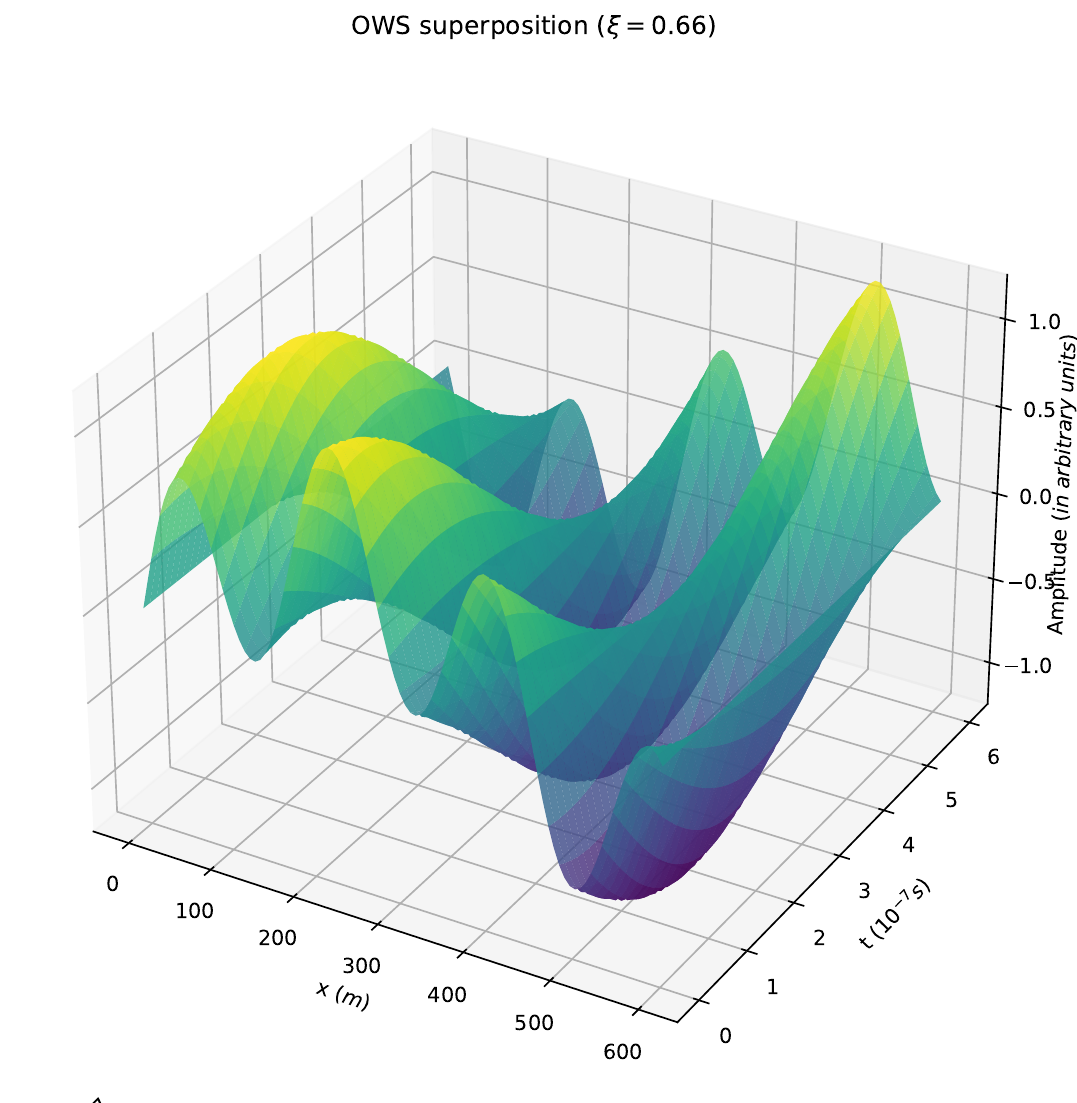}}
\centerline{\includegraphics[width=2.0in]{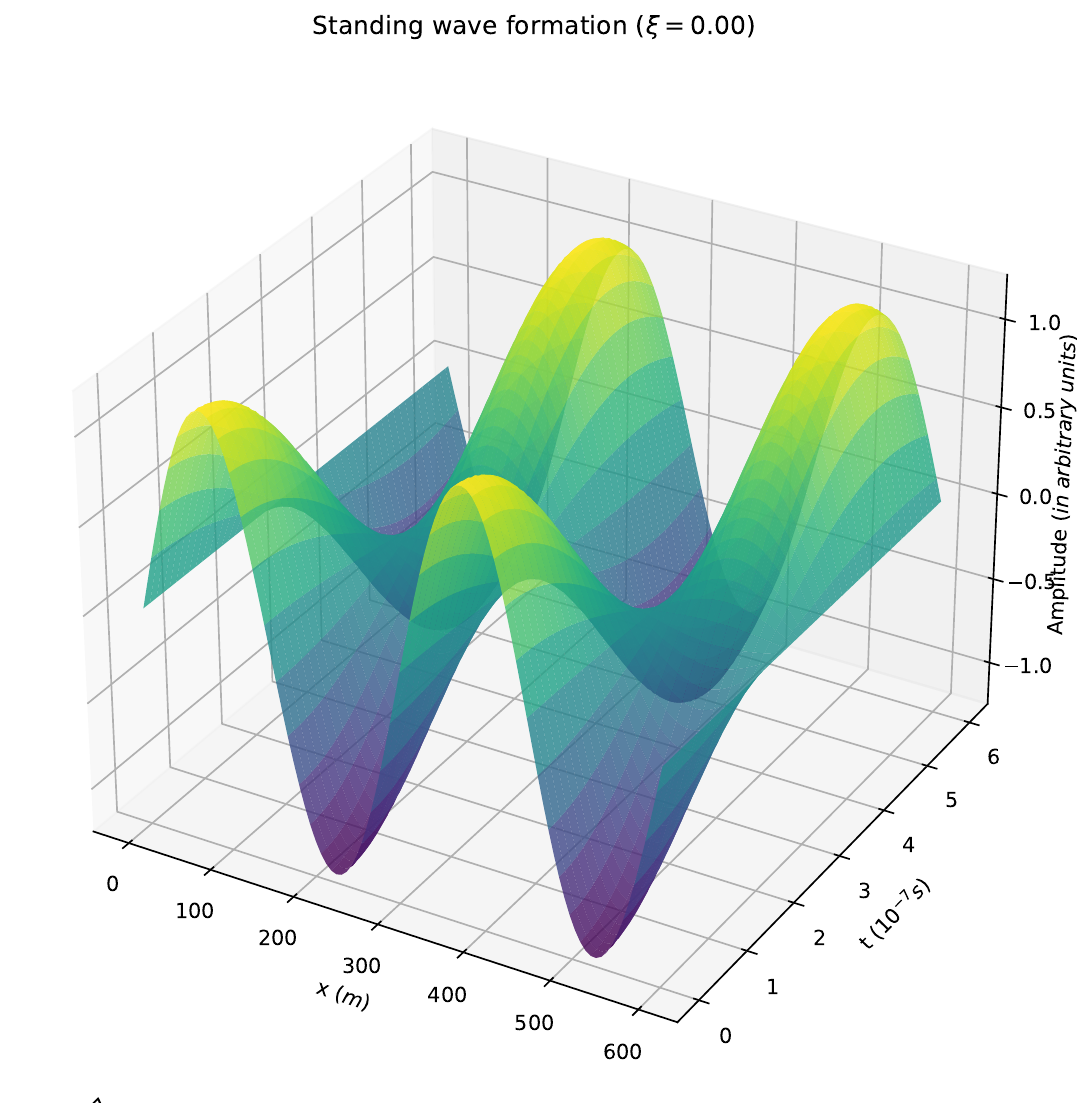}}

%  \caption{}
%  \label{fig:sub-third}
%\end{figure}
%\begin{figure}
%  \centering
  % include fourth image
%  \includegraphics{stn-6.pdf}  
\vspace*{8pt}
  \caption{ The continuous time variation of standard standing wave ($\xi = 0.00$) (upper panel) and the resultant wave obtained by applying the AVS gauge transformation to the standing wave with $\xi=0.66$ (lower panel).}
  \label{fig:sub-second}
\end{figure}	

%\caption{The discrete time variation of a light signal in radio wave length ($\lambda=300$ m) forming (a) a standard standing wave (b) the standing wave under AVS gauge transformation with $\xi=0.66$. The continuous time variation of (c) standard standing wave ($\xi = 0.00$) and (d) standard standing wave under AVS gauge transformation with $\xi=0.66$.}
%\label{fig:fig}
%\end{figure}

%%%%%%%%%%%%%%%%%%%%%%%%%%%%%%%%%%%%%%%%

%The issue is now whether one can experimentally detect standing wave, particularly the location of anti-nodes without any choice of synchrony.  

\section{Experimental evidence of standing waves:}

The issue is now whether one can experimentally detect standing wave, particularly the location of anti-nodes without any choice of synchrony.  
The nature of standing waves including the precise position of the antinodes has already been detected experimentally. Historically, optical standing waves were first detected experimentally by Wiener \cite{wiener1890stehende}. He put a thin photographic emulsion to a silver mirror surface and illuminated the mirror normally by a quasi-monochromatic light from a sodium arc lamp. Several blackened equidistant parallel bands were observed on the emulsion after development. The blackened maxima were due to the intersection between the emulsion and the anti-nodes of the standing waves. Subsequently, many experiments unambiguously confirmed the existence of specific nodes and antinodes \cite{pan2017direct,cha2010demonstration}. Similar experiments with sound wave are even more abundant.

\section{Conclusion}

In conclusion, the present analysis suggests that the standing waves are never formed in the superposition of incident and reflected waves unless we fix the synchronization gauge so that the one-way speed of the signals in two directions is the same. For general non-vanishing synchronization gauge parameter the resultant wave formed in a superposition of incident and reflected waves will be characterized by zero amplitude at locations $x = 0, \lambda/2, \lambda, \, ...,$ similar to the nodes of standing waves, and the resulting wave travels between two minima. There will be no fixed anti-nodes, the location of the maxima depends on the AVS synchronization gauge. Such behavior is not consistent with the observations. 
 
A true gauge field should not affect any physical observable. The gauge characteristics of the AVS synchronization field is thus questionable. The AVS synchronization field originates from the invariance of physical laws under a coordinate transformation. However, such coordinate transformations are not as innocuous as they may appear at first sight. For instance, the magnitude of $\xi$ needs to be less than 1, otherwise, the velocity will become negative. So $\xi$ is not arbitrary. %, rather it may be restricted by various physical consideration. Note that 
In fact there is absolutely no reason why speed of light in one direction should be different from that in another direction. Malament argued that the Einstein synchronization is uniquely definable from the causal structure (the causal connectivity of events) of Minkowski space-time \cite{Malament77}. Malament reached his conclusion assuming the causal automorphism, i.e. all pairs of world points on an orthogonal hyperplane are mapped to another orthogonal hyperplane as well as mapped to itself (though several researchers found the stated assumptions of causal automorphism to a near-tautology \cite{Anderson1977, Redhead93, Grunbaum01}).

The present analysis suggests that Einstein's synchronization is the only acceptable synchronization technique. This finding has many implications. For instance, if the conventionality issue becomes non-existent, the distant simultaneity becomes a law of nature itself and Lorentz transformation remains the only acceptable relativistic transformation leaving aside various others proposed. The present finding eliminates the (unphysical) synchronization gauge freedom of special relativity.   
\\
%\section*{Declaration on conflict of interest}
%The authors declare that they do not have any conflicts of interest. 
%We declare that we don't have any potential competing or financianl/personal interests in submitting the manuscript.  
\ack{
%\section*{Acknowledgement}
The authors thank the anonymous reviewers for the comments that helped to correct and improve the manuscript. BR likes to thank IUCAA (Pune, India) for their hospitality during his stay there under the visiting associateship programme.}

\section*{Data availability statement} The present work is a theoretical work and it has no associated data.

\section*{References}
%\bibby{

\end{document}